\documentstyle[epsf,twocolumn,prl,aps]{revtex}
\begin{document}

\title{Pulsed sputtering during homoepitaxial surface growth:
layer--by--layer forever}

\author{Joachim Jacobsen and James P. Sethna}

\address{Laboratory of Atomic and Solid State Physics, Cornell
  University, Ithaca, NY 14853-2501}

\date{\today}

\maketitle

\begin{abstract}
The homoepitaxial growth of initially flat surfaces has so far
always led to surfaces which become rougher and rougher as the number
of layers increases: even in systems exhibiting ``layer by layer'' growth
the registry of the layers is gradually lost.  We propose that
pulsed glancing--angle sputtering, once per monolayer, can in principle lead to
layer--by--layer growth that continues indefinitely, if one additional
parameter is controlled.  We illustrate our suggestion with a fairly realistic
simulation of the growth of a Pt (111) surface, coupled with a 
simplified model for the sputtering process.
\end{abstract}

\pacs{PACS numbers: 81.15.-z,
64.60.Qb
}

When depositing atoms on a surface, one often wants control of the 
growth morphology.  At the most primitive level, when depositing atom X
on a low--index, flat surface of the X crystal, it would be nice to 
be able to ensure that the resulting surface remained flat!  

This simple goal has often been difficult to achieve in practice, especially
in metals.  At low
temperatures, one observes three-dimensional growth: inter--layer
mobility is low, and the second layer starts growing as soon as
the first layer gains any substantial coverage.  At intermediate temperatures,
one observes what is called two-dimensional or layer--by--layer growth.
Layer--by--layer is used to describe systems with oscillations in some
measured property (antiphase scattering of He\cite{Rosenfeld}, 
RHEED\cite{RHEED}, low-energy electrons~\cite{henzler},
or X-rays\cite{Cooper}), 
decaying slowly as the number of layers increases.  For a few materials
under special conditions, this decay can be quite slow\cite{Platinum}, but
under most circumstances it decays over a few tens of layers.
At higher temperatures, for slightly mis-cut surfaces, one can have
a step--flow regime which typically exhibits quite stable layered growth.

In this paper, we consider the question of how one might achieve indefinite
layer--by--layer growth: a mode-locked state\cite{StrogatzMPAF}
where the surface irregularities
due to the growth process would remain bounded and oscillations in the 
properties would continue forever.  Such long--range order despite the noise
(random fluctuations in deposition, nucleation, 
and growth) would be separated from the traditional decaying layer--by--layer
growth by a phase transition.
The key is to {\it periodically force
the system, pulsing in synchrony with the deposition of each monolayer}
(keeping in phase by using one of the real--time oscillatory measurements
described above).  Pulsing the temperature, pulsing the
deposition rate, and pulsing with an ion--assisted anneal have been used
to good effect experimentally.\cite{Rosenfeld}  We argue that pulsing 
these quantities {\it cannot} lead to indefinite flat growth, but that pulsed
sputtering can yield layer--by--layer growth forever.

\begin{figure}[thb]
  \begin{center}
    \leavevmode
    \epsfxsize=8cm
    \epsffile{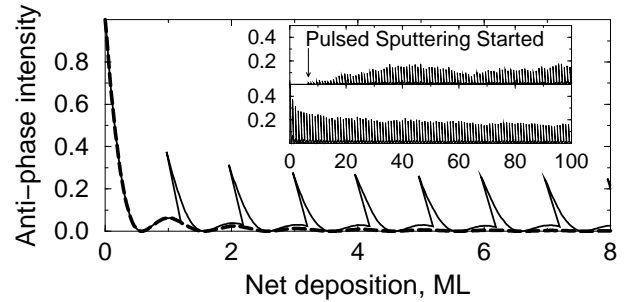}
  \end{center}
\caption{Layer by layer growth, with and without pulsed sputtering.
The main plot shows the simulated antiphase intensity $I$;
the oscillations with a traditional
thermal deposition (dashed line) decay rapidly. 
Pulsed sputtering of $\lambda =$ 25\%  
of a monolayer is done after each deposition of 1.25  monolayers, yielding
sharp jumps in the curves.  (Here the sputtering phase $\mu=0.20$.)
The lower curve in the inset shows the peaks of the jumps for
the first hundred monolayers deposited: we expect the layers to stay flat
indefinitely, for the correct choice of $\mu$.  The upper inset curve
is the signal from an initially rough sample prepared by first depositing
several layers without sputtering. When the sputter--deposit sequence is
started, we observe that the surface regains a flat interface.
}
\label{fig:oscillations}
\end{figure}

The dashed line in figure~\ref{fig:oscillations} shows a typical thermal
growth on a surface.  It is a numerical simulation of Pt/Pt(111) at 130 K,
grown at one monolayer/second, with parameters determined 
using Effective Medium Theory and available experimental
information as described in \cite{ptmodel}.  
The figure shows 
$ I = \left( \sum_{i=0}^{\infty} (-1)^i (\theta_{i+1}-\theta_i)\right)^2$ as a
function of time, where
$\theta_i$ is the fractional coverage in the $i$th layer; this measure
corresponds closely to  
what is measured in antiphase scattering probes used in the experimental
systems.  Notice that the signal vanishes whenever the number of atoms
in odd and even layers become equal.
Notice that the peaks decay as more layers are deposited: as the
surface begins to span several layers, the surface morphologies
at integer and non-integer monolayer coverages become indistinguishable.
One must note that, for the parameters we simulate,
the surface does not grow wildly rough even for thermal deposition,
where we observe four oscillations in $I$.

This decaying layered growth is usually described by theories focusing
on the nucleation and growth of islands on the surface.\cite{layered}
Most of the pulsed
attempts to improve the stability of layered growth\cite{Rosenfeld} have
been motivated by the nucleation and growth theories, and have
deliberately increased the nucleation rate at the onset of a new layer
(where depositing on top of existing islands is not a concern), while
reducing it thereafter.

There is another school of theoretical models, which focus not on
individual islands but rather on the effect of fluctuations in the
deposition rate and the role of diffusion within a continuum description
for the height of the surface.  These models predict that the random
fluctuations in deposition will always overwhelm the available diffusive
mechanisms for retaining a flat interface, on sufficiently long length
and time scales.\cite{continuum} 
There are several mechanisms and models\cite{rough} for this diffusive
smoothing of surfaces, but {\it they all predict that the surfaces will
eventually become rough:} the noise is independent of wavelength, and
the diffusion becomes feeble at long wavelengths.  However unlikely it
is to nucleate on top of an existing island, diffusion cannot transport
the extra atoms from one region of the surface to another fast enough:
eventually the extra atoms in one region will nucleate extra layers.
Pulsing the temperature, pulsing the deposition rate, or pulsed
annealing with an ion beam only changes the effective diffusion rates on
the surface, and does not fundamentally alter this conclusion: we need a
non--diffusive mechanism.

How can we smooth the surface in a more effective way?  High--energy
atomic beams incident at glancing angles to the surface are a known way
of generating flat surfaces\cite{labanda}: the beam preferentially
sputters atoms off the mountains and hills.  Especially for groups using
energetic beams for growing surfaces\cite{Cooper}, it would seem natural
to try to use a pulsed sputtering mechanism.  Consider starting with a
flat surface with an initial deposition of $1+\mu$ monolayers.  We then
repeatedly sputter off $\lambda$ and deposit $1+\lambda$ monolayers, so
as to always start the sputtering at an integer plus $\mu$ monolayers
coverage. If a surplus of atoms is deposited onto a region of the flat
surface, there will be a surplus of adatoms at the time of sputtering,
and thus the sputtering will remove extra mass from the region.  This
non--diffusive mechanism, being independent of wavelength, should in
principle be able to produce indefinite layered growth.\cite{Unpulsed}

We decided to test these ideas with a simple model.  Building on our
well--characterized,\cite{ptmodel}
physically realistic solid--on--solid kinetic Monte--Carlo
model for thermal deposition and growth on Pt(111), we implemented a primitive,
simplistic model for glancing--angle rotating--beam 
sputtering.\cite{grooved} We wanted atoms in an intact
layer to be immune from sputtering, solitary atoms and edges of small islands
to be sputtered at high rates, and pit edges to be at least partially
shielded.  We sputter equally from
each of the six directions lying along atomic rows.  An atom is immune from
sputtering from a given direction if it is shadowed by another atom upwind
in the same row and the same monolayer, within a distance $L=5$ interatomic
distances.  Figure~\ref{fig:oscillations} shows that this model can produce
smooth growth, seemingly forever.

This sputtering model we imagine might correspond roughly to a beam at
an angle $\arctan(1/L)=11^\circ$. Our model is likely over--optimistic in
that intact monolayers and shadowed atoms are protected completely.
It is using a pessimistic angle of attack: angles from $1^\circ$ to
$15^\circ$ have been used for smoothing rough surfaces\cite{labanda}, and
a more glancing angle in our simulation would produce much smoother surfaces.
This kind of simplistic model
is particularly useful in studies of the qualitative features
and feasibility of a new method, and indeed it immediately uncovered an
important issue neglected in our discussions so far. 

\begin{figure}[thb]
  \begin{center}
    \leavevmode
    \epsfxsize=8cm
    \epsffile{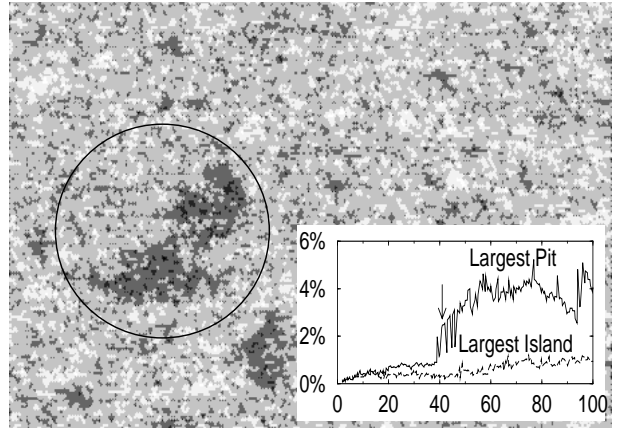}
  \end{center}
\caption{{\bf Critical pit.}  This is a snapshot of our simulation
    with $\lambda = 0.25$, and $\mu = 0.1$, after 41 layers have been
    deposited. Lower layers are darker grey (pits).  The inset shows
    a plot of the percentage of the system size occupied by the largest
    island and largest pit, as a function of how many monolayers
    have been deposited.  The arrow in the inset
    shows the time of the snapshot.  Until 39 layers, all we see are small
    pits and islands which appear and disappear.  At this point, we see
    a large pit form (circled), which grows irregularly thereafter.
    The oscillations in the anti-phase intensity
    (as in figure~\ref{fig:oscillations}) die out slowly after the
    pit nucleates.
}
\label{fig:criticalpit}
\end{figure}

Figure~\ref{fig:criticalpit} shows a surface grown with our sputtering
schedule with a different value of $\mu$, after growing 41 monolayers.
Notice the pit. Up until this point in the simulation, we saw excellent
layer--by--layer growth, similar to that shown in
figure~\ref{fig:oscillations}; subsequent to this frame the oscillations
die away.  We interpret this behavior as the nucleation of a critical
pit, analogous to critical droplets at first--order phase
transitions.\cite{Langer}

Consider what happens to a pit of radius $R_n$ under a cycle of depositing
$1+\lambda$ monolayers and sputtering off $\lambda$. During the deposition,
atoms falling on the upper layer will typically nucleate new islands, which
grow and merge to raise the height by one.  Atoms landing inside the pit
will stick to its outer edges; the pit will fill in (given a large 
Ehrlich--Schwoebel barrier) at about the time a whole monolayer is deposited.
However, the region of the former pit will remain depressed, since new
islands start nucleating and growing only after it fills in.   As
a zeroth approximation, a new pit one level higher of the same radius
will exist after one cycle of deposition and sputtering: $R_{n+1} \sim R_n$.

Consider a flat step on the surface ---
interpretable either as a pit or an island of infinite radius.  Under
one cycle of deposition and sputtering, there is no reason
to expect that the attachment at the step edge will balance the sputtering.
(That is, our model has no symmetry between pits and islands.)
One expects the step edge to move by a distance $\Delta$, where we define
positive $\Delta$ to represent the growth of the pit (lower terrace).
To a first approximation for large radius, we expect the
pit area after one cycle on average to change: $R_{n+1} = R_n + \Delta$.

Now, since the
edges of small pits are partially shielded, and islands are more exposed
to the sputtering, the net effect after one cycle is to remove less material
in existing pits, and more near existing islands. The smaller the pit radius
$R_n$, the more protected is the pit, and the higher is the net deposition
after the entire cycle.  To a second approximation, we expect that the pit
will be re-formed at a new radius $R_{n+1} = R_n + \Delta - \Sigma/R_n$.
The term $-\Sigma/R_n$ represents the physics of the self--shielding
for small pits which tends to make them shrink; it also makes small islands
shrink (negative radius).  This is the term found for detachment--limited
coarsening for islands on surfaces\cite{BadriMcLean},
is the first term in a Taylor series in the curvature of the
edge, and can be derived with a simple geometrical argument based
on shielded sites.

\begin{figure}[thb]
  \begin{center}
    \leavevmode
    \epsfxsize=9cm
    \epsffile{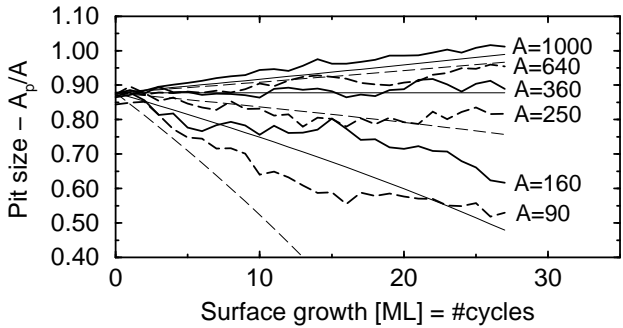}
  \end{center}
  \vspace*{-25mm}
\caption{Large pits grow, small pits decay, for $\lambda = 0.25$, and
$\mu=0.15$.  Shown is the evolution of the area $A_p$ of a pit for
6 different initial pit sizes $A$, each 10\%
of the initial area of the simulation, averaged over 100 runs each
(except for size 100, with 85 runs).
We follow the evolution of the pit by measuring its size every time
there is an integer number of monolayers down.
The thin curves are equation~\ref{eq:pitsize} fit to the data 
(if $a$ is the lattice spacing, then $\Delta \sim 0.09 a$, and 
$\Sigma \sim 0.9 a^2$, with an initial shrinkage of the island radius
of about 6\%.) The critical pit size $R_c$ is estimated to be $280\pm60$ in
area, or about ten lattice constants in radius.
}
\label{fig:biggrow}
\end{figure}

Finally, there are the stochastic fluctuation in the pit radius. We
expected that the dominant source of fluctuations would be the
fluctuations in the number of atoms deposited within the area of the
pit, which should scale as the square root of its area: hence the radius
fluctuations will be independent of radius.
\begin{equation}
  \label{eq:pitsize}
  R_{n+1} = R_n + \Delta - \Sigma/R_n + \Omega\, \xi_n.
\end{equation}
$\Omega\, \xi_n$ in equation~\ref{eq:pitsize} is the noise term:
$\xi_n$ is a random variable with mean zero and standard deviation one, and
$\Omega$ gives the strength of the noise.
Measuring these fluctuations directly, in the
interesting range $0<\mu<0.2$, we've verified that they are indeed roughly
independent of $R$ and of width $\Omega \sim 0.65a$.
Figure~\ref{fig:biggrow} shows the results of a fit of the average shrinking
and growing of pits, using equation~\ref{eq:pitsize} without noise
($\Omega=0$).

The layer--by--layer growth in our model ends when the fluctuating noise
produces a pit of the critical size
\begin{equation}
  \label{eq:criticalpit}
  \langle R_c \rangle = \Sigma/\Delta,
\end{equation}
after which the pit grows by itself to macroscopic size.  Our
equation~\ref{eq:pitsize} can be thought of as a thermal random--walk in
radius, with step size $\Omega$, temperature $T=2 \Omega^2$, and potential
$\Sigma \log(R) - (R-1)\Delta$:\cite{StrangePotential} the critical radius
is the local maximum $V_{max}$ in the potential.  One can solve a continuum
approximation to equation~\ref{eq:pitsize} for the rate of formation of
large pits, per density of pits of size $R=1$:
\begin{eqnarray}
  \label{eq:current}
J &= \Delta (2 \Delta/\Omega^2)^{2 \Sigma/\Omega^2}/
   \Gamma[1 + 2\Sigma/\Omega^2, 2 \Delta/\Omega^2]\hskip 0.5truein \\
  =&
  \Delta \left\{(2 \Sigma/(e \Omega^2))^{2 \Sigma/\Omega^2}/
   \Gamma[1 + 2 \Sigma/\Omega^2, 2 \Delta/\Omega^2]\right\}
e^{-V_{max}/T},\nonumber
\end{eqnarray}
where $\Gamma$ is the incomplete gamma function.

The last of equations~\ref{eq:current} shows the connection with
traditional critical droplet theory.\cite{Langer,BadriMcLean} Here the
term in curly brackets is a prefactor, $\Delta$ is a bound on the
velocity at which one could cross the barrier, and $e^{-V_{max}/T}$ is
the Boltzmann probability of sitting at the critical radius.

How can we grow layer--by--layer forever?  Clearly, {\it we wish to set 
$\Delta$ to zero, imposing a long--wavelength symmetry between islands and pits}
sending the nucleation rate $J$ to zero.
All three constants $\Delta$, $\Sigma$, and $\Omega$ in
equation~(\ref{eq:pitsize})
will depend on temperature, deposition rate, sputtering angle, other
adsorbates on the surface, the
fraction $\lambda$ sputtered, and the point $\mu$ during the deposition
of a monolayer that the sputtering occurs.  If by varying any of these
parameters we can set $\Delta=0$ without making $\Sigma<0$, we ought to
suppress the nucleation altogether, and sustain layered growth indefinitely.

\begin{figure}[thb]
  \begin{center}
    \leavevmode
    \epsfxsize=8cm
    \epsffile{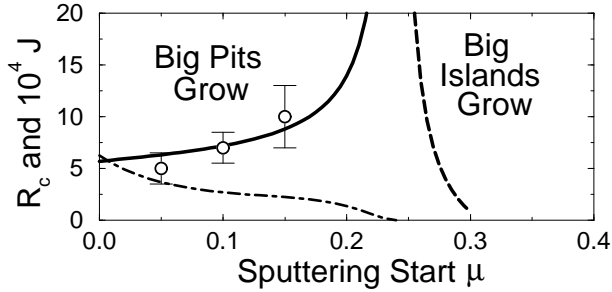}
  \end{center}
\caption{Critical pit size $R_c$ (in units of the lattice constant $a$) 
and large pit formation rate $J$ (in units of nucleated large pits per
monolayer per vacancy) as a function of sputtering
phase $\mu$.  The points with error bars are eyeball estimates from plots
like figure~\ref{fig:biggrow}.
The upper curves are from equation~\ref{eq:criticalpit} with sinusoidal
fits to $\Delta(\mu)$ and $\Sigma(\mu)$.
Because $\Delta$ changes sign at $\mu_c \sim 0.24$, the critical pit size
diverges there; for $\mu<\mu_c$ large pits are unstable (solid curve), and
for $\mu > \mu_c$ large islands become unstable (dashed curve).  When the
critical pit size diverges, the rate of large pit formation $J$ goes to zero
(dot--dashed curve).
}
\label{fig:divergence}
\end{figure}

Fitting to simulations like those shown above in
figure~\ref{fig:biggrow}, we
have measured the critical pit size $R_c$, $\Delta$, and $\Sigma$ as functions 
of $\mu$.  We find $\Delta$, $\Sigma$, and an initial island shrinkage all fit 
well to the form $a \sin(2 \pi \mu + \phi) + b$.  
Direct measurements of the critical pit size show a divergence
where our sinusoidal interpolation for $\Delta$ changes sign
(figure~\ref{fig:divergence}).\cite{Sigmachange}
Above $\mu_c$, where $\Delta < 0$, we expect all pits to be stable and
large islands to be unstable.\cite{IslandProblems}

Can we show that the surfaces remain flat near $\mu_c$?
The sputtered simulation shown in figure~\ref{fig:oscillations} was 
done at $\mu = 0.20$ (near $\mu_c=0.24$ where the lifetime of flat growth
diverges); the inset shows that the oscillations are persisting as long as we 
have simulated.  Equation~(\ref{eq:current}) predicts the rate
of formation of large pits $J(\mu=0.20)$ to be  $1.4\times 10^{-4}$ times
the density of pits of size one.
For values of $\mu$ far from $\mu_c$ the oscillations decay rapidly:
at $\mu=0.7$ the oscillations die roughly as they do for a thermal
growth without sputtering (although the rms roughness for the sputtered
surface is much smaller than that for a thermally grown, unsputtered
surface even away from $\mu_c$).
Figure~\ref{fig:divergence} also shows our prediction of the nucleation
rate $J$ for large pits: proportional to the inverse of the lifetime for flat
growth.

Finally, in perhaps the most convincing demonstration that our method
is working, note back in figure~\ref{fig:oscillations} the second plot
in the inset.  It shows a rough surface becoming flat
once we start our pulsed sputtering schedule!

In trying to grow flat layers, why not go for perfection?  We argue
that the methods used heretofore to grow flat surfaces (pulsing 
temperature, deposition rate, ion--assisted diffusion, ...) cannot compete 
in the end with the stochastic noise in the deposition rate.
We claim that pulsed sputtering, smoothing once per deposited monolayer,
can in principle yield layer--by--layer growth oscillations
that last forever, provided that one parameter is tuned to a critical
value.

The authors wish to thank Barbara Cooper, Randy Headrick, 
Karsten Jacobsen, and Kamal Bhattacharya for inspiration and helpful
conversations.  This
work was supported  by  the National  Science Foundation  through  the
Cornell Materials Science Center  NSF-DMR-9121654.


\begin{references}

\bibitem{Rosenfeld} G. Rosenfeld, N.~N. Lipkin, W.~Wulfhekel, J.~Kliewer,
K.~Morgenstern, B.~Poelsema, and G.~Comsa, {\sl Appl. Phys. A} {\bf 61},
455 (1995).

\bibitem{RHEED} T. Sakamoto, T. Kawamura, S. Nago, G. Hashiguchi, K. Sakamoto,
K. Kuniyoshi, {\sl J. Cryst. Growth}, {\bf 81}, 59 (1987).

\bibitem{henzler} M. Henzler, {\sl Surf. Sci.} {\bf 298}, 369 (1993). 

\bibitem{Cooper} ``X-Ray Scattering Study of the Surface Morphology of
Au (111) During Ar$^+$ Ion Irradiation'', M.~V. Ramana Murty, T.~Curcic,
A.~Judy, B.~H. Cooper, A.~R. Woll, J.~D. Brock, S. Kycia, and R.~L. Headrick,
preprint.

\bibitem{Platinum} Silicon up to 2200 periods (T. Sakamoto {\it et al.},
{\sl Appl. Phys. Lett.} {\bf 47}, 617 (1985)), platinum (111) over
150 oscillations (R. Kunkel, B. Poelsema, L. K. Verheij, and G. Comsa, 
{\sl Phys. Rev. Lett.} {\bf 65}, 733 (1990), R. Kunkel,
{\sl J\"ul-Bericht 2526} (Forschungszentrum J\"ulich, J\"ulich,
1991, p. 94).  It's likely that the layer-by-layer persistence is related
to the reconstructions on the terraces which do not extend to the islands,
yielding two different diffusion rates (T. Michely, M. Hohage, S. Esch,
and G. Comsa, ``An Origin of Undamped Oscillations in Homoepitaxial Growth'',
preprint.)  We emphasize that real platinum would not grow layer--by--layer 
at our simulation temperature of 130K, nor (as it happens) does our
effective-medium simulation reconstruct at 650K.

\bibitem{StrogatzMPAF} Renato E. Mirollo and Steven H. Strogatz, 
{\sl SIAM Journal on Applied Mathematics}, {\bf 50}, 1645 (1990), 
L. Balents and M. P. A. Fisher, {\sl Phys. Rev. Lett.} {\bf 75}, 4270 (1995).

\bibitem{Unpulsed} Unpulsed sputtering could also yield flat growth.

\bibitem{ptmodel} J. Jacobsen, K.~W. Jacobsen, P. Stoltze,
  and J.~K. N\o rskov, {\sl Phys. Rev. Lett.} {\bf 74}, 2295 (1995),
  J.~Jacobsen, K.~W.~Jacobsen, and J.~K.~N\o rskov, {\sl Scanning
  Microscopy}, in print.

\bibitem{layered} Atoms deposited
on top of existing islands can either nucleate into new islands (leading
to three--dimensional growth) or can attach to the edge of the island
(after crossing the Ehrlich--Schwoebel energy barrier at the perimeter).
Competition between these two rates leads to a transition between
layered and three-dimensional growth, with island density and island
size being important parameters: see
J. Tersoff, A. Denier van der Gorn, and R. M. Tromp,
{\sl Phys. Rev. Lett.} {\bf 72}, 266 (1994).

\bibitem{continuum}
Decompose the height $h({\bf x})$ in Fourier space
$h_{\bf k}$.  The random noise introduced by depositing a monolayer will
increase the mean square of each Fourier component $\langle h_{\bf
k}^2\rangle$ by the same amount $\eta$ (the Fourier transform of random
noise is flat).  On the other hand, the various diffusion processes on
the surface will tend to flatten the surface. For example,
above the roughening transition an initial sinusoidal perturbation
will decay exponentially with a rate given by the inverse fourth power
of the wavelength (W. W. Mullins, {\sl J. Appl. Phys.} {\bf 28}, 333 (1957),
C. Herring, {\sl J. Appl. Phys.} {\bf 21}, 301 (1950)).
This result has been seen below the roughening transition (M. E. Keeffe,
C. C. Umbach, and J. M. Blakely, {\sl J. Phys. Chem. Solids (UK)}, {\bf 55},
965 (1994), S. Tanaka, C. C. Umbach, J. M. Blakely, R. M. Tromp, and M. Mankos,
{\sl J. Vac. Sci. Technol. A}, {\bf 15 pt.2}, 1345 (1997), and 
simulation: M.~V. Ramana Murty and B.~H. Cooper, {\sl Phys. Rev. B}
{\bf 54}, 10377 (1996)), perhaps because of small miscuts or perhaps
because of crossover effects.  Thus in this regime $d \langle h_{\bf
k}^2\rangle / dt \propto - k^4 \langle h_{\bf k}^2\rangle + \eta$,
yielding a stationary state whose roughness grows as wavevector
shrinks, as $k^{-4}$.  

\bibitem{rough}  Continuum theories of growth below the roughening transition:
M. Kardar, G. Parisi, and Y. C. Zhang, {\sl Phys. Rev. Lett.} {\bf 56}, 889
(1986),
S. F. Edwards and D. R. Wilkinson, {\sl Proc. R. Soc. London, Ser. A} {\bf 381},
17 (1982),
J. Villain, {\sl J. Phys. (France) I} {\bf 1}, 19 (1991),
Z. W. Lai and S. Das Sarma, {\sl Phys. Rev. Lett.} {\bf 66}, 2348 (1991),
J. A. Stroscio, D. T. Pierce, M. Stiles, A. Zangwill, and L. M. Sander,
{\sl Phys. Rev. Lett.} {\bf 75}, 4246 (1995).

\bibitem{grooved} We got grooves when sputtering from one direction only.

\bibitem{labanda} J.~G.~C. Labanda, S.~A. Barnett, and L. Hultman,
{\sl Appl. Phys. Lett.} {\bf 66}, 3114 (1995) cleaved at $15^\circ$;
M. Wissing, M. Holzworth, D.~S. Simeonova, and K.~J. Snowdon, 
{\sl Rev. Sci. Instrum.} {\bf 67}, 4314 (1996) used $5-10^\circ$;
M. Holzwarth, M. Wissing, D.~S. Simeonova, S. Tzanev, K.~J. Snowdon, and
O.~I. Yordanov, {\sl Surf. Sci.} {\bf 331-333}, 1093 (1995) used
$3-5^\circ$; U.~von~Gemmingen and R. Sizmann, {\sl Surf. Sci.} {\bf 114},
445 (1982) sputtered at $1^\circ$.

\bibitem{Langer} J. S. Langer, {\sl Ann. Phys.} {\bf 41} 108 (1967);
{\bf 54} 258 (1969).

\bibitem{BadriMcLean} B. Krishnamachari, J.G. McLean, B.H. Cooper, 
and J.P. Sethna, {\sl Phys. Rev. B} {\bf 54}, 8899 (1996);
J.G. McLean, B. Krishnamachari, D.R. Peale, E. Chason, 
J.P. Sethna, and B.H. Cooper, {\sl Phys. Rev. B} {\bf 55}, 1811 (1997). 

\bibitem{StrangePotential}
The effective noise is much higher for our non-equilibrium islands than it is
for thermal pits: hence power laws rather than 
exponentials in equation~(\ref{eq:current}).

\bibitem{Sigmachange} Above about $\mu=0.3$ $\Sigma<0$; our theory no longer
applies.

\bibitem{IslandProblems} Our exploration of this region 
does indeed show islands substantially larger than the corresponding pits,
and we qualitatively saw large island clusters nucleate and destroy
the flatness.  The large shadowing length $L$ led to diffuse islands which
whose sizes were hard to measure.

\end{references}
\end{document}